# Synthesis and Structural Characterization of Highly Tetragonal (1-x)Bi(Zn$_{1/2}$Ti$_{1/2}$)O$_3$-xPbTiO$_3$ Piezoceramics

Jyoti Sharma, Rishikesh Pandey and Akhilesh Kumar Singh*

*School of Materials Science & Technology, Indian Institute of Technology, Banaras Hindu University, Varanasi- 221005, India.*
*Email: akhilesh_bhu@yahoo.com

**Abstract.** We present here the comprehensive X-ray diffraction (XRD) and dielectric measurement of (1-x)Bi(Zn$_{1/2}$Ti$_{1/2}$)O$_3$-xPbTiO$_3$ (BZT-xPT) piezoceramics with x=0.65, 0.70, 0.75 and 0.80. Powder X-ray diffraction data reveals the tetragonal structure (space group P4*mm*) of BZT-xPT ceramics for all the compositions.

## INTRODUCTION

Recently, there is great interest in PbTiO$_3$ based morphotropic phase boundary (MPB) systems such as Pb(Zr$_x$Ti$_{1-x}$)O$_3$ (PZT), (1-x)Pb(Mg$_{1/3}$Nb$_{2/3}$)O$_3$-xPbTiO$_3$ (PMN-xPT) etc having excellent piezoelectric property and wide industrial application for devices such as sensors, actuators and transducers.[1] In the search of such valuable systems, very recently a new Bi based compound Bi(Zn$_{1/2}$Ti$_{1/2}$)O$_3$ (BZT) has been investigated with very high, calculated ionic polarization ~ 138 μC/cm$^2$ and large tetragonality (c/a ~ 1.211), which is much greater than PbTiO$_3$ (c/a ~ 1.064).[2] However, some limitations with the synthesis of BZT is that high pressure is required to synthesize the pure perovskite phase. Formation of non perovskite phases start when it is heated above 550 $^0$C at ambient pressure. Recently, several investigations have been carried out aiming to get morphotropic phase boundary by solid solution formation of BZT with ferroelectric BaTiO$_3$, Na$_{0.5}$K$_{0.5}$NbO$_3$ and PbTiO$_3$ etc. The Solid solution formation of BZT with PbTiO$_3$ increases the tetragonality of PbTiO$_3$ but, surprisingly solid solution formation of BZT with BaTiO$_3$, Na$_{0.5}$K$_{0.5}$NbO$_3$ diminishes the long range ferroelectric order and leads to cubic structure with increasing BZT concentration.[3-5] In the present work we have explored the (1-x)Bi(Zn$_{1/2}$Ti$_{1/2}$)O$_3$-xPbTiO$_3$ solid solution with the objective of stabilization of the perovskite phase of BZT.

## EXPERIMENTAL DETAILS

BZT-xPT solid solution with compositions x=0.65, 0.70, 0.75 and 0.80 were prepared by conventional solid state route. Stoichiometric amount of analytical reagent (AR) grade Bi$_2$O$_3$ (HIMEDIA, ≥99.5%), ZnO (HIMEDIA, 99%), TiO$_2$ (HIMEDIA, ≥99%), PbCO$_3$ (100 %) were mixed with acetone for 6 h by ball milling. The mixed powder was then calcined in a muffle furnace at 800 $^0$C for 6 h. The calcined powder was checked for phase purity by using an 18 kW rotating anode Cu-based Rigaku x-ray powder diffractometer. This calcined powder was mixed with binder (2% polyvinyl alcohol water solution) to form the pellets using a cylindrical die (12 mm diameter) and uniaxial hydraulic press. These pellets were heated at 500 $^0$C for 10 h to burnout the binders. The pellets were sintered at 1000 $^0$C for 3 h. Each composition was sintered in Bi$_2$O$_3$ and PbO atmosphere in a closed alumina crucible to avoid their evaporation during sintering at high temperature. The frequency dependent dielectric constant and loss tangent (tanδ) were measured by using Nova Control (ALPHA A) high performance frequency analyzer.

## CRYSTAL STRUCTURE OF BZT-xPT

Selected X-ray diffraction (XRD) pattern of BZT-xPT ceramics with x=0.65, 0.70, 0.75 and 0.80 is shown in Fig. 1. A close examination of XRD profiles shown in Fig. 1 suggest that the structure of BZT-xPT is tetragonal, since there is distinguishable splitting for

(200) and (220) peaks for all the compositions. As BZT content increases, splitting in (200)/(002) profiles increases. It is expected due to localized strain at domain boundaries and growth of 'c' lattice parameter as reported earlier by Suchomel *et al.* for BZT.[3] To confirm the tetragonal structure of different compositions of BZT-xPT, we carried out full pattern analysis of the XRD data with tetragonal structure (P4*mm* space group) using Rietveld profile refinement (Fig. 2 (a), (b) and (c)) with x=0.65, 0.75 and 0.80 by FULLPROF package.[6] A very good fit between observed and calculated profiles is obtained for each composition except for few week reflections which correspond to the impurity phase identified as $ZnTiO_3$. The peaks of this impurity phase are marked with asterisks in Fig. 2. This confirms that the structure of BZT-xPT is tetragonal with the space group P4*mm*. We have plotted the tetragonality as well as lattice parameters (Fig. 3) which shows that tetragonality increases as the BZT contain is increased.

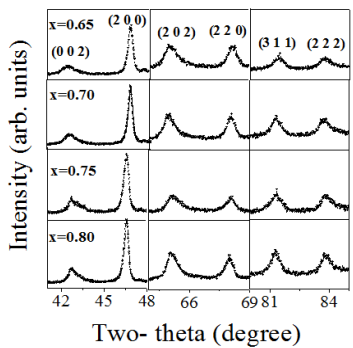

**FIGURE 1.** Selected XRD profiles of BZT-xPT ceramics.

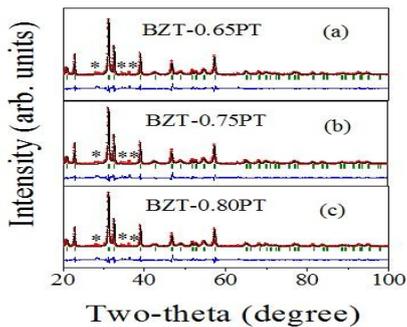

**FIGURE 2.** Rietveld fits of powder XRD patterns for BZT-xPT ceramics using tetragonal (P4*mm*) structure.

## DIELECTRIC STUDY

Fig. 4 shows the composition dependence of dielectric permittivity ($\varepsilon'$) and loss tangent (tan$\delta$) for BZT-xPT compositions. It is observed that dielectric permittivity is almost constant with varying the compositions. Since the structure of BZT-xPT remains tetragonal for these compositions, such variations of dielectric constant are expected. The loss tangent (tan$\delta$) shows slightly increasing trend with decreasing PT concentration.

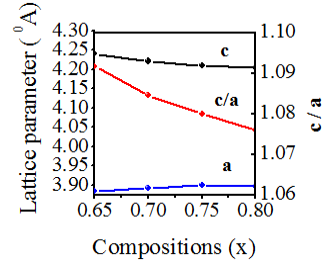

**FIGURE 3.** Variation of lattice parameters and tetragonality for BZT-xPT ceramics.

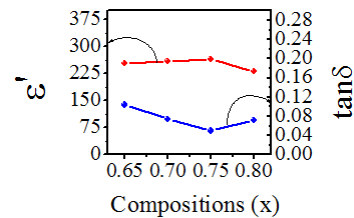

**FIGURE 4.** Variation of dielectric permittivity ($\varepsilon'$) and loss tangent (tan$\delta$) for BZT-xPT ceramics.

## SUMMARY

Nearly phase pure BZT-xPT ceramics were prepared in single step calcination and sintering. The small impurity phase is identified as $ZnTiO_3$. The structure of all the compositions studied is tetragonal (P4*mm*). The dielectric permittivity of BZT-xPT ceramics is nearly composition independent in the composition range 0.65≤x≤0.80.

## ACKNOWLEDGMENTS

Rishikesh Pandey acknowledges University Grant Commission (UGC), India for the financial support as Junior Research Fellowship (JRF).